\documentclass[referee]{raa}            

\usepackage{graphicx,times}             
\usepackage{natbib}
\usepackage{amssymb,amsmath}
\bibpunct{(}{)}{;}{a}{}{,}

\usepackage[a4paper=true,dvipdfm=true,pagebackref=true]{hyperref}
\hypersetup{colorlinks = true, linkcolor = green, anchorcolor = red, citecolor = blue, filecolor = red, pagecolor = red, urlcolor = red}

\begin{document}

   \title{The first photometric analysis of the totally eclipsing contact binary V811 Cep
}

   \volnopage{Vol.0 (20xx) No.0, 000--000}      
   \setcounter{page}{1}          

   \author{Xiang Gao
      \inst{1}
   \and Kai Li
      \inst{1}
   \and Xing Gao
      \inst{2}
   \and Yuan Liu
      \inst{3}
   }

   \institute{Shandong Provincial Key Laboratory of Optical Astronomy and Solar-Terrestrial Environment, Institute of Space Sciences, Shandong University, Weihai, 264209, China  ({\it e-mail: kaili@sdu.edu.cn (KL)})\\
        \and
             Xinjiang Astronomical Observatory, Urumqi 830011, China\\
        \and
             Qilu Institute of Technology, Jinan 250200, China\\
             \vs\no
   {\small Received; accepted}}

\abstract{ The first photometric analysis of V811 Cep was carried out. The first complete light curves of V, R and I bands are given. The analysis was carried out by Wilson-Devinney (W-D) program, and the results show that V811 Cep is a median-contact binary ($f=33.9(\pm4.9)\%$) with a mass ratio of 0.285. It is a W-subtype contact binary, that is, the component with less mass is hotter than the component with more mass, and the light curves are asymmetric (O'Connell effect), which can be explained by the existence of a hot spot on the component with less mass. The orbital inclination is $i=88.3^{\circ}$, indicating that it is a totally eclipsing binary, so the parameters obtained are reliable. Through the O-C analyzing, it is found that the orbital period decreases at the rate of $\dot{P}=-3.90(\pm0.06)\times 10^{-7}d \cdot yr^{-1}$, which indicates that the mass transfer occurs from the more massive component to the less massive one.
}

   \authorrunning{Gao et al. }            
   \titlerunning{The First Photometric Analysis of V811 Cep }  

   \maketitle

%
%
\section{INTRODUCTION}           
\label{sect:intro}

W UMa type binaries belong to contact binaries, which are considered to be a
universal type of eclipsing binaries. Both components are full of or beyond their respective Roche lobes and share common envelopes (\citealt{Lucy+L+B+1968a}). The light curve of W UMa type binaries is generally EW type, which shows that the two eclipses are roughly equal (\citealt{Qian+etal+2017}). So we know that these two components have similar surface temperatures, but at the same time they have very different masses (\citealt{Lucy+L+B+1968b}). Binnendijk divided W UMa type binaries into A-subtype and W-subtype. The temperature of the more massive component of A-subtype is higher than that of the less massive component, while W-subtype is on the opposite (\citealt{Binnendijk+L+1970}). At present, there is no clear conclusion about the formation and evolution of contact binaries. Therefore, it is very necessary to observe and study W UMa type contact binaries as many as possible.

As an important physical parameter of contact binaries, the mass ratio is sometimes obtained only by
light curve analysis, which is inaccurate in some cases. According to the spectroscopic and photometric
mass ratios of 80 contact binaries collected by \citet{Pribulla+etal+2003}, partially eclipsing binaries'
data shows some deviations, while totally eclipsing binaries' data matches well. This shows that the mass ratio obtained by photometric analysis is accurate and feasible for totally eclipsing binaries. \citet{Terrell+D+Wilson+R+E+2005} confirmed this conclusion.

Sometimes, the two maxima in the light curve of eclipsing binaries are not equal, which is called the O'Connell effect (\citealt{O'Connell+D+J+K+1951}). Generally speaking, there are the following explanations: starspots due to magnetic activity (\citealt{Binnendijk+L+1964}); the material accretion between two components (\citealt{Shaw+J+S+1994}); the circumstellar material surrounding the binary (\citealt{Liu+Yang+2003}); material asymmetry caused by coriolis force (\citealt{Zhou+Leung+1997}), etc. Among them, the existence of starspots is usually thought to be the reason of the O'Connell effect (\citealt{Qian+etal+2013}; \citealt{Li+2014+AJ}; \citealt{Zhou+etal+2016}).

V811 Cep was identified as a W UMa type eclipsing binary by \citet{Sokolovsky+Chekhovich+Korotkiy}. They observed the target and determined that the period was 0.250762 days. The light curve of V811 Cep exhibits that it is a totally eclipsing binary system. No one has yet studied its light curve and orbital period variation. Therefore, we did the first research on V811 Cep.


\section{OBSERVATIONS}
\label{sect:Obs}

\begin{figure}
\begin{center}
\includegraphics[width=80mm]{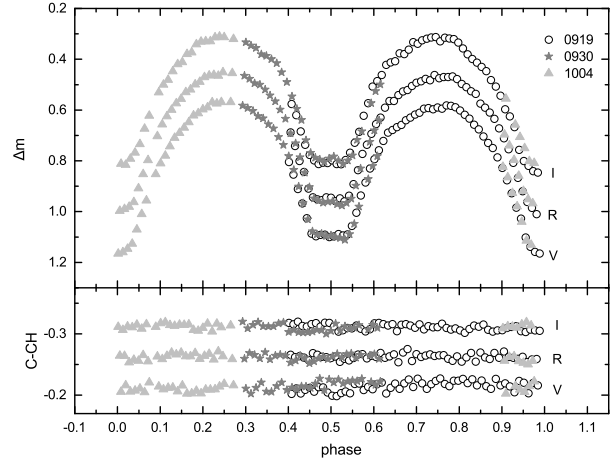}
\end{center}
\caption{CCD photometric light curves of V811 Cep in the VRI bands obtained using NEXT. Different symbols represent different days.}\label{Fig1}
\end{figure}

Photometric observations of V811 Cep were carried out by the Xingming Observatory on the nights of September 19, 30 and October 4, 2018, using the Ningbo Bureau of Education and Xinjiang Observatory Telescope (NEXT) with an aperture of 60 cm. The NEXT uses a back-illuminated FLI 230-42 CCD camera. The camera with a format of $2048 \times 2048$ pixels provides a field of view about $22' \times 22'$. We used the standard Johnson-Cousin-Bessel UBVRI filter system in the observation. All valid images were processed by the C-Munipack\nolinebreak\footnotemark[1] \footnotetext[1]{http://c-munipack.sourceforge.net/} program. C-Munipack is a CCD photometric data processing software based on Munipack. It supports Windows and Linux operating systems and supports graphical user interface and command line modes. All images were corrected for bias and flat. In order to reduce the CCD data, the methods of aperture photometry and differential photometry were adopted. Then, we determined the magnitude differences between the variable and comparison stars, and between the comparison and check stars. The two stars 2MASS 20042618+6106309 and 2MASS 20042943+6103411 were chosen as the comparison star and check star. Their informations are shown in Table~\ref{VCCH1}, which "V811 Cep" is the eclipsing binary, "C" is the comparison star and "CH" is the check star. This table lists the names, coordinates and magnitudes of these stars. We used the following equation to calculate the orbital phases,
\begin{equation}
\label{E1}
Min.I(HJD)=2458396.11006+0.^{d} 250762\times E
\end{equation}
Figure~\ref{Fig1} depicts the VRI light curves of V811 Cep. The uncertainties of the light curves of V, R and I bands are 0.007 mag, 0.005 mag, 0.005 mag, respectively. As depicted in the figure, the brightness varies continuously and the two minima are almost as deep. The light curve obviously shows the characteristics of EW type. The flat secondary eclipse shows that V811 Cep is a totally eclipsing binary, and the duration time of flat eclipse is about 26.65 min. From the figure, we can find the obvious O'Connell effect, which is reflected in that the maximum of phase 0.75 in V band is about 0.021 mag brighter than the maximum of phase 0.25. We have determined three minimum moments according to the observations.

\begin{table}
\begin{center}
\caption[Table1]{ Details of the V811 Cep and the Comparison and Check Stars}\label{VCCH1}
\resizebox{\textwidth}{!}{
 \begin{tabular}{lcccccc}
  \hline\noalign{\smallskip}
Star &  Names      & RA & Dec & J(mag) & H(mag) & K(mag)                    \\
  \hline\noalign{\smallskip}
V811 Cep      & 2MASS 20041682+6105323&$20^{h}04^{m}17^{s}$&$+61^{o}05^{'}32^{''}$&12.707&12.236&12.120  \\
Comparison (C)& 2MASS 20042618+6106309&$20^{h}04^{m}26^{s}$&$+61^{o}05^{'}31^{''}$&12.374&12.050&11.975  \\
Check (CH)    & 2MASS 20042943+6103411&$20^{h}04^{m}29^{s}$&$+61^{o}05^{'}41^{''}$&12.470&12.032&11.996  \\
  \noalign{\smallskip}\hline
\end{tabular}}
\end{center}
\end{table}

\section{ORBITAL PERIOD INVESTIGATIONS}
\label{sect:data}

\begin{figure}
\begin{center}
\includegraphics[width=80mm]{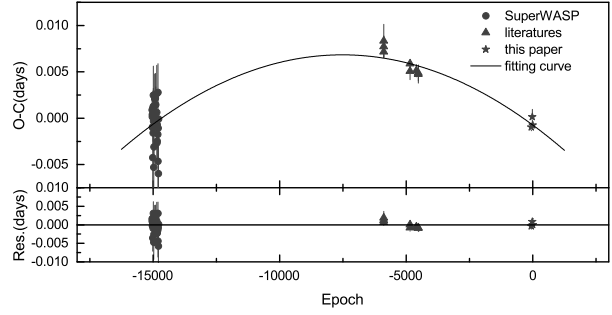}
\end{center}
\caption{This figure depicts the O-C values of V811 Cep. The upper black solid line represents the fitting curve of Equation(2). The residuals are shown at the bottom of the figure.}\label{Fig2}
\end{figure}

We identified a total of 45 minimum moments for V811 Cep. Except for 3 minimum moments obtained from our observations, 34 minimum moments were obtained from the SuperWASP (calculated by us) and 8 from the \citet{Jurysek+J}. All the data are CCD observations with a time span from 2005 to 2018. All the minimum moments are listed in Table~\ref{VCCH2}. Using Equation (1), the O-C values were computed, the corresponding O-C curve is depicted in Figure~\ref{Fig2}. It can be seen from the upper part of the Figure~\ref{Fig2} that the O-C curve shows a downward parabola trend, indicating that the orbital period is reduced. Using the least square method, we can obtained the following equation:
\begin{eqnarray}
\label{E2}
Min.I& = 2458396.109318(\pm0.000050)\\\nonumber
& + 0.250759986(\pm0.000000028)\times E\\
& - 1.34(\pm0.02)\times 10^{-10} \times E^{2}.\nonumber
\end{eqnarray}
The rms of the fit of Equation (2) is 0.00199. If linear fitting is used, the rms is 0.00277, which is bigger than that of the quadratic fit.
In Equation (2), the coefficient of quadratic term reveals that the decreasing rate of the period is $dP/dt=3.90(\pm0.06)\times 10^{-7}d \cdot yr^{-1}$.The black line in the upper part of Figure~\ref{Fig2} shows the long-term period decrease. After the downward parabola fitting was removed, the residuals are shown in the lower part of Figure~\ref{Fig2}. However, we don't confirm that the O-C diagram show an obvious downward parabola in Figure~\ref{Fig2}, because the minima time are concentrated in three clusters of points and there are two big gaps in the curves. More minima data are needed to confirm that.

\begin{table}
\begin{center}
\caption[Table2]{Eclipsing Times for V811 Cep }\label{VCCH2}
 \begin{tabular}{ccccrrrc}
  \hline\noalign{\smallskip}
HJD & Error & Method & Min & E & O-C & Residual & Reference \\
  \hline\noalign{\smallskip}
54625.59910 &	0.00092 &	CCD &	p &	$-$15036.0  &	   0.00089 & 	 0.00166 &	(1)  \\
54626.60166 &	0.00125 &	CCD &	p &	$-$15032.0  &	   0.00039 & 	 0.00114 &	(1)  \\
54627.60448 &	0.00080 &	CCD &	p &	$-$15028.0  &	   0.00014 & 	 0.00089 &	(1)  \\
54628.60640 &	0.00071 &	CCD &	p &	$-$15024.0  &	$-$0.00100 &  $-$0.00026 &	(1)  \\
54629.60952 &	0.00102 &	CCD &	p &	$-$15020.0  &	$-$0.00094 &  $-$0.00021 &	(1)  \\
54630.60926 &	0.00093 &	CCD &	p &	$-$15016.0  &	$-$0.00426 &  $-$0.00354 &	(1)  \\
54631.61699 &	0.00106 &	CCD &	p &	$-$15012.0  &	   0.00040 & 	 0.00112 &	(1)  \\
54632.62076 &	0.00144 &	CCD &	p &	$-$15008.0  &	   0.00111 & 	 0.00182 &	(1)  \\
54635.62725 &	0.00098 &	CCD &	p &	$-$14996.0  &	$-$0.00159 &  $-$0.00090 &	(1)  \\
54636.63317 &	0.00085 &	CCD &	p &	$-$14992.0  &	   0.00127 & 	 0.00195 &	(1)  \\
54637.63743 &	0.00135 &	CCD &	p &	$-$14988.0  &	   0.00247 & 	 0.00314 &	(1)  \\
54638.63770 &	0.00133 &	CCD &	p &	$-$14984.0  &	$-$0.00033 & 	 0.00033 &	(1)  \\
54639.64186 &	0.00102 &	CCD &	p &	$-$14980.0  &	   0.00077 & 	 0.00142 &	(1)  \\
54641.64192 &	0.00138 &	CCD &	p &	$-$14972.0  &	$-$0.00529 &  $-$0.00466 &	(1)  \\
54642.64718 &	0.00069 &	CCD &	p &	$-$14968.0  &	$-$0.00310 &  $-$0.00247 &	(1)  \\
54650.55149 &	0.00113 &	CCD &	p &	$-$14936.5  &	   0.00210 & 	 0.00266 &	(1)  \\
54652.55596 &	0.00089 &	CCD &	p &	$-$14928.5  &	   0.00044 & 	 0.00099 &	(1)  \\
54656.56993 &	0.00095 &	CCD &	p &	$-$14912.5  &	   0.00216 & 	 0.00268 &	(1)  \\
54657.57064 &	0.00059 &	CCD &	p &	$-$14908.5  &	$-$0.00019 & 	 0.00031 &	(1)  \\
54661.58453 &	0.00138 &	CCD &	p &	$-$14892.5  &	   0.00145 & 	 0.00192 &	(1)  \\
54670.48431 &	0.00066 &	CCD &	p &	$-$14857.0  &	$-$0.00095 &  $-$0.00055 &	(1)  \\
54670.61330 &	0.00117 &	CCD &	s &	$-$14856.5  &	   0.00265 & 	 0.00306 &	(1)  \\
54671.48812 &	0.00087 &	CCD &	p &	$-$14853.0  &	$-$0.00021 & 	 0.00019 &	(1)  \\
54671.61434 &	0.00114 &	CCD &	s &	$-$14852.5  &	   0.00063 & 	 0.00103 &	(1)  \\
54672.49050 &	0.00054 &	CCD &	p &	$-$14849.0  &	$-$0.00089 &  $-$0.00050 &	(1)  \\
54673.49179 &	0.00066 &	CCD &	p &	$-$14845.0  &	$-$0.00266 &  $-$0.00228 &	(1)  \\
54674.49504 &	0.00095 &	CCD &	p &	$-$14841.0  &	$-$0.00248 &  $-$0.00210 &	(1)  \\
54674.62222 &	0.00112 &	CCD &	s &	$-$14840.5  &	$-$0.00068 &  $-$0.00031 &	(1)  \\
54680.51414 &	0.00088 &	CCD &	p &	$-$14817.0  &	$-$0.00175 &  $-$0.00143 &	(1)  \\
54682.52480 &	0.00040 &	CCD &	p &	$-$14809.0  &	   0.00278 & 	 0.00309 &	(1)  \\
54683.52044 &	0.00115 &	CCD &	p &	$-$14805.0  &	$-$0.00464 &  $-$0.00434 &	(1)  \\
54684.52708 &	0.00103 &	CCD &	p &	$-$14801.0  &	$-$0.00106 &  $-$0.00077 &	(1)  \\
54686.52830 &	0.00082 &	CCD &	p &	$-$14793.0  &	$-$0.00597 &  $-$0.00569 &	(1)  \\
54688.54030 &	0.00083 &	CCD &	p &	$-$14785.0  &	$-$0.00009 & 	 0.00017 &	(1)  \\
56919.35950 &	0.00030 &	CCD &	p &	$-$5889.0   &      0.00772 & 	 0.00116 &	(2)  \\
56919.48550 &	0.00040 &	CCD &	p &	$-$5888.5   &      0.00834 & 	 0.00178 &	(2)  \\
56919.60970 &	0.00050 &	CCD &	p &	$-$5888.0   &      0.00716 & 	 0.00060 &	(2)  \\
57180.40390 &	0.00030 &	CCD &	p &	$-$4848.0   &      0.00506 &  $-$0.00090 &	(2)  \\
57180.53010 &	0.00040 &	CCD &	p &	$-$4847.5   &      0.00588 &  $-$0.00008 &	(2)  \\
57241.46540 &	0.00040 &	CCD &	p &	$-$4604.5   &      0.00512 &  $-$0.00066 &	(2)  \\
57261.40090 &	0.00040 &	CCD &	p &	$-$4525.0   &      0.00475 &  $-$0.00097 &	(2)  \\
57261.52640 &	0.00020 &	CCD &	p &	$-$4524.5   &      0.00487 &  $-$0.00085 &	(2)  \\
58381.18928 &	0.00030 &	CCD &	p &	$-$59.5  	&   $-$0.00096 &  $-$0.00040 &	(3)  \\
58392.22408 &	0.00027 &	CCD &	p &	$-$15.5  	&      0.00015 & 	 0.00080 &	(3)  \\
58396.11006 &	0.00018 &	CCD &	p &	   0.0      &   $-$0.00074 &  $-$0.00005 &	(3)  \\

  \noalign{\smallskip}\hline
\end{tabular}
\end{center}
\tablecomments{0.86\textwidth}{(1) This paper (SuperWASP); (2) \citet{Jurysek+J}; (3) This paper.}
\end{table}

\begin{table}
\begin{center}
\caption[Table3]{ Photometric Solutions of V811 Cep }\label{VCCH3}
{
 \begin{tabular}{lcccccc}
  \hline\noalign{\smallskip}
          & Photometric Elements  &       & Photometric Elements &       & Photometric Elements&      \\
Parameter & No Spot               & Error & With Dark Spot       & Error & With Hot Spot       & Error\\
  \hline\noalign{\smallskip}

g$_{1}$=g$_{2}$ & 0.32 & Assumed & 0.32 & Assumed & 0.32 & Assumed \\
A$_{1}$=A$_{2}$ & 0.5 & Assumed & 0.5 & Assumed & 0.5 & Assumed \\
T$_{1}$(K) & 5264 & Assumed & 5264 & Assumed & 5264 & Assumed \\
T$_{2}$(K) & 5275 & $\pm12$ & 5327 & $\pm13$ & 5289 & $\pm9$ \\
q($M_{2}/M_{1}$) & 0.293 & $\pm0.003$ & 0.297 & $\pm0.003$ & 0.285 & $\pm0.003$ \\
i($^{\circ}$) & 89.7 & $\pm0.6$ & 89.3 & $\pm0.1$ & 88.3 & $\pm0.4$ \\
$\Omega_{in}$ & 2.451 & ... & 2.460 & ... & 2.433 & ... \\
$\Omega_{out}$ & 2.268 & ... & 2.274 & ... & 2.255 & ... \\
$\Omega_{1}=\Omega_{2}$ & 2.388 & $\pm0.008$ & 2.399 & $\pm0.011$ & 2.372 & $\pm0.009$ \\
$L_{1}/(L_{1}+L_{2})(V)$ & 0.743 & $\pm0.001$ & 0.731 & $\pm0.001$ & 0.746 & $\pm0.001$ \\
$L_{1}/(L_{1}+L_{2})(R)$ & 0.743 & $\pm0.001$ & 0.733 & $\pm0.001$ & 0.747 & $\pm0.001$ \\
$L_{1}/(L_{1}+L_{2})(I)$ & 0.744 & $\pm0.001$ & 0.735 & $\pm0.001$ & 0.747 & $\pm0.001$ \\
r$_{1}$(pole) & 0.474 & $\pm0.001$ & 0.470 & $\pm0.002$ & 0.472 & $\pm0.001$ \\
r$_{1}$(side) & 0.514 & $\pm0.002$ & 0.509 & $\pm0.002$ & 0.512 & $\pm0.002$ \\
r$_{1}$(back) & 0.554 & $\pm0.002$ & 0.539 & $\pm0.002$ & 0.541 & $\pm0.002$ \\
r$_{2}$(pole) & 0.268 & $\pm0.004$ & 0.274 & $\pm0.005$ & 0.272 & $\pm0.004$ \\
r$_{2}$(side) & 0.281 & $\pm0.005$ & 0.287 & $\pm0.006$ & 0.286 & $\pm0.005$ \\
r$_{2}$(back) & 0.322 & $\pm0.010$ & 0.332 & $\pm0.013$ & 0.332 & $\pm0.011$ \\
$f$ & 34.3$\%$ & $\pm4.5\%$ & 32.6$\%$ & $\pm5.7\%$ & 33.9$\%$ & $\pm4.9\%$ \\
$\theta(^{\circ})$ & ... & ... & 44.58 & $\pm3.17$ & 111.46 & $\pm3.61$ \\
$\psi(^{\circ})$ & ... & ... & 48.07 & $\pm6.70$ & 27.41 & $\pm3.36$ \\
$r(^{\circ})$ & ... & ... & 13.99 & $\pm0.76$ & 18.25 & $\pm0.83$ \\
T$_{f}$ & ... & ... & 0.759 & $\pm0.036$ & 1.208 & $\pm0.014$ \\
$\sum(O-C)_{i}^{2}$ & 0.0025 & ... & 0.0022 & ... & 0.0020 & ... \\

  \noalign{\smallskip}\hline
\end{tabular}}
\end{center}
\end{table}

\section{LIGHT CURVE SOLUTIONS}

We used the 2013 version of the W-D program (\citealt{Wilson+R+E+1979}, \citealt{Wilson+R+E+1990}; \citealt{Wilson+R+E+Devinney+E+J}) to analyze the light curves of V811 Cep. As mentioned earlier, V811 Cep is a totally eclipsing binary, so the results of photometric solutions will be very reliable. The surface temperature of the primary component was estimated by the calculated color index. In the calculation, the reddening and extinction (\citealt{Schlafly+Finkbeiner}) were taken into account. After calculating the temperature corresponding to the g-i, B-V, and J-K, the average value is 5264K. According to \citet{Lucy+L+B+1967} and \citet{Rucinski+S+M}, for stars with convective envelopes, the gravity darkening and bolometric albedo coefficients were set to be g$_{1,2}$ = 0.32 and $A_{1,2}=0.5$, respectively. As for the square root bolometric and bandpass limb-darkening coefficients, they were taken from \citet{van+Hamme+W}. We found that the solutions converge well in mode 3, and the adjustable parameters were the orbital inclination, $i$, the average surface temperature of the secondary Star 2, $T_{2}$, the dimensionless potential, ($\Omega_{1}=\Omega_{2}$ in mode 3), and the monochromatic luminosity of Star 1, $L_{1}$.

\begin{figure}
\begin{center}
\includegraphics[width=80mm]{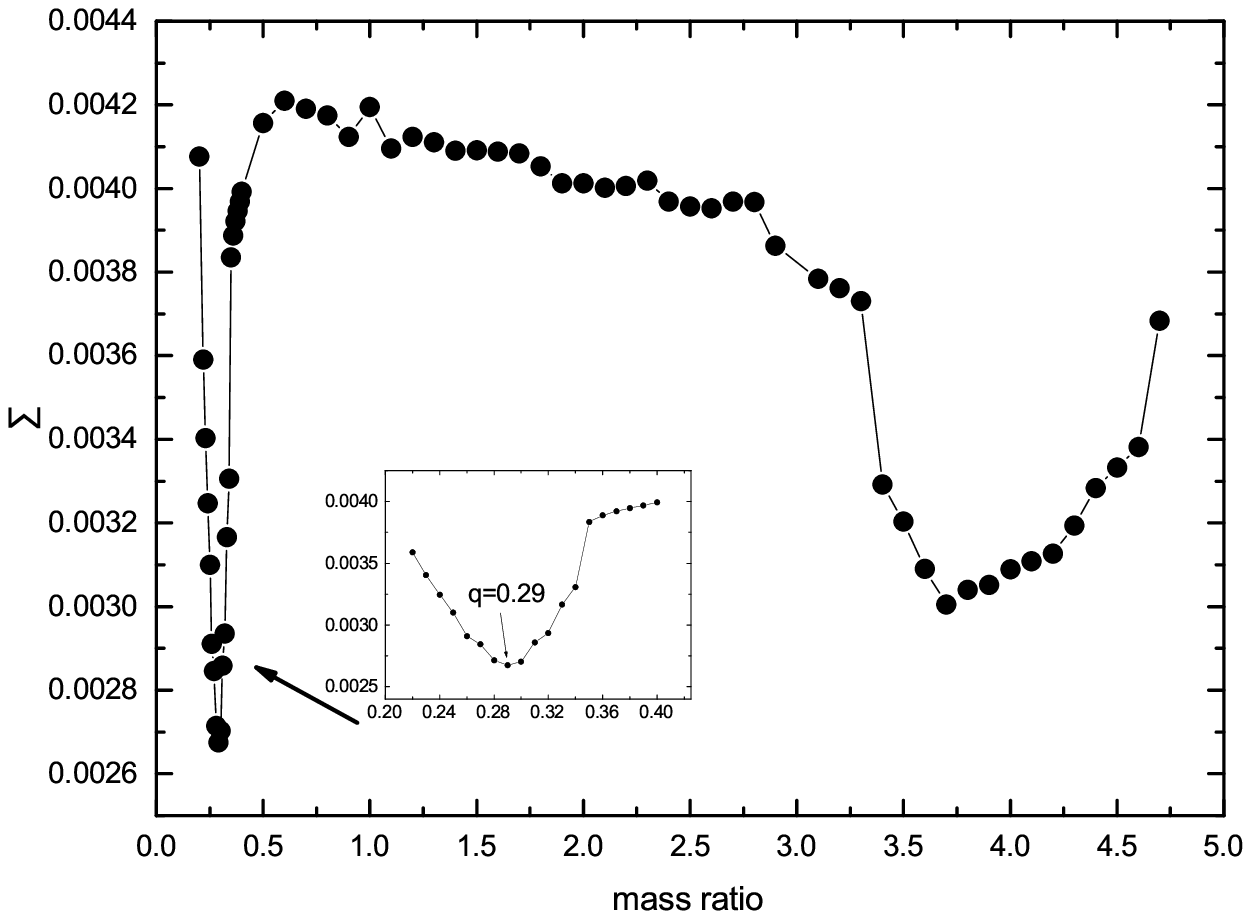}
\end{center}
\caption{This figure depicts the sum square of residuals $\sum$ and mass ratio of V811 Cep.}\label{Fig3}
\end{figure}

\begin{figure}
\begin{center}
\includegraphics[width=80mm]{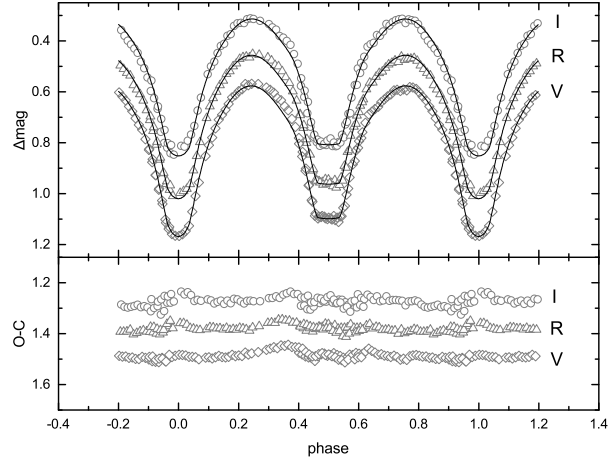}
\end{center}
\caption{This figure displays theoretical light curves calculated without taking into account the spots. The hollow circles, triangles and diamonds represent the light curve in the I, R and V bands, respectively. The values of O-C are depicted in the bottom.}\label{Fig4}
\end{figure}

Since there were no investigations of V811 Cep, the mass ratio was determined by the q-search method. The solutions of the mass ratio from 0.2 to 4.7 were searched, and the relationship between the sum of the square $\Sigma$ of the residuals and the mass ratio $q$ is shown in Figure~\ref{Fig3}. Two minima were found in the figure, at the mass ratios of $q$ = 0.29 and $q$ = 3.7, respectively. However, as depicted in Figure~\ref{Fig3}, the minimum of $q$ = 0.29 is much lower than that of $q$ = 3.7. Hence we chose $q$ = 0.29 as the initial value of $q$ and took $q$ as an adjustable parameter. After differential corrections, the finial photometric solutions are displayed in Table~\ref{VCCH3} and the theoretical light curve is depicted in Figure~\ref{Fig4}. Where the definition of the contact degree is $f=(\Omega_{1}-\Omega)/(\Omega_{1}-\Omega_{2})$. All the errors in Table~\ref{VCCH3} are not real errors, but are due to the fitting errors calculated by the W-D program, which are underestimated (\citealt{Prasa+A+Zwitter+T+2005}). In our solution, V811 Cep shows the characteristics of W-type contact binaries. 
The physical parameters obtained from this light curve solution using a q-search determined mass ratio are only preliminary, not the final reliable results. Therefore, further spectroscopic observations of V811 Cep are needed to confirm these results.

\begin{figure}
\begin{center}
\includegraphics[width=80mm]{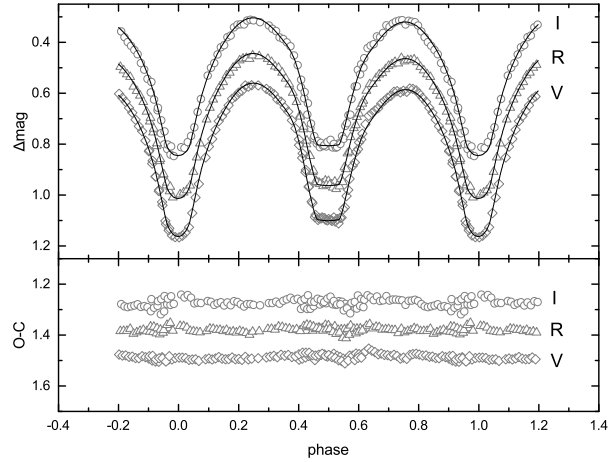}
\end{center}
\caption{This figure displays observed (symbols) and theoretical (solid lines) light curves calculated with a hot spot on the secondary component of V811 Cep. The meanings of the symbols are the same as those of Figure~\ref{Fig4}. The values of O-C are depicted in the bottom.}\label{Fig5}
\end{figure}

However, in the process of solving, the theoretical light curves near the two maxima do not agree well with the observed values. Because the primary and secondary stars are fast-rotating solar-like stars, they should have similar activity characteristics to the sun, such as starspots (e.g., \citealt{Qian+etal+2015}; \citealt{Zhou+etal+2015}; \citealt{Li+2018+pasp}). Therefore, in order to reasonably explain the asymmetry of the light curve, the spot model was introduced. There are four parameters of spot model in W-D program, which are spot central latitude ($\theta$), spot central longitude ($\phi$), spot angular radius ($r$) and spot temperature factor ($T_{f}=T_{d}/T_{0}$, where $T_{f}$ is the ratio of spot temperature $T_{d}$ to star surface temperature $T_{0}$). Cool spot and hot spot were introduced into the primary and secondary components, respectively. As shown in Table~\ref{VCCH3}, the residual of the latter is smaller, so the fitting result of hot spot is better. The final fitting light curves are shown in Figure~\ref{Fig5}.

\section{DISCUSSION AND CONCLUSIONS}
\label{sect:discussion and conclusions}
According to photometric solutions, we concluded that V811 Cep is a contact binary system with a degree of contact of $f=33.9\%$. The mass ratio was finally determined to be $q=0.285$. It is concluded that the primary star is a K0-type star with the temperature of 5264K (\citealt{Pecaut+Mamajek+2013}). The temperature difference between the two components is 25K, this system is in thermal contact (\citealt{Li+2019+mnras}). The orbital inclination of the system is 88.3$(\pm0.4)^{\circ}$, indicating that the system is a totally eclipsing system, so the physical parameters obtained are reliable (\citealt{Pribulla+etal+2003}; \citealt{Terrell+D+Wilson+R+E+2005}). The O'Connell effect is interpreted as the existence of a hot spot on the less massive component. The temperature of the hot spot is about 1100K higher than the surface temperature of the secondary star, and its area covers about 2.5$\%$ of the entire photosphere surface.

According to \citet{Li+2019+RAA}, the absolute parameters can be estimated based on the distance from the $Gaia$ mission (\citealt{Gaia+Brown+etal}). First of all, according to the  distance $D$ and $M_{V}=m_{V}-5logD+5-A_{V}$, the absolute magnitude of V811 Cep can be obtained, where $m_{V}$ is the visual magnitude of the V band (\citealt{Samus+2017}), $D$ represents the distance which is given by the $Gaia$ mission (\citealt{Bailer+etal+2018}) and $A_{V}$ is the extinction value obtained from \citet{Schlafly+Finkbeiner}. Then, the total luminosity of the system was calculated by equations $M_{bol}=-2.5logL/L_{\odot}+4.74$ and $M_{bol}=M_{V}+BC_{V}$ ($M_{bol}$ is the absolute bolometric magnitude and $BC_{V}$ is the bolometric correction from \citealt{Pecaut+Mamajek+2013}), and then through the relationship between $L_{1}$ and $L_{1}+L_{2}$ given in Table~\ref{VCCH3}, the luminosity of each component ($L_{1}$ and $L_{2}$) can be obtained respectively. Next, assuming black-body radiation, the radius of each star can be determined by $L=4\pi\sigma T^{4}R^{2}$, and the semi-major axis $a$ of the binary system can be obtained from the absolute radius and relative radius of each star (the value of $a$ was the average). Finally, from the Kepler's third law $M_{1}+M_{2}=0.0134a^{3}/P^{2}$ and the mass ratio $q$, we got the mass of each component. From the above steps, we obtained the absolute parameters of V811 Cep. The absolute parameters and the parameters required in the calculation are listed in Table~\ref{VCCH4}.

\begin{table}
\begin{center}
\caption[Table4]{ Absolute Parameters of V811 Cep }\label{VCCH4}
\resizebox{\textwidth}{!}{
 \begin{tabular}{ccccccccccccc}
  \hline\noalign{\smallskip}
Parameters&D&$V_{max}$&$M_{V}$&$BC_{V}$&$M_{bol}$&$L_{1}$&$L_{2}$&$R_{1}$&$R_{2}$&$a$&$M_{1}$&$M_{2}$\\
&(pc)&(mag)&(mag)&(mag)&(mag)&($L_{\odot}$)&($L_{\odot}$)&($R_{\odot}$)&($R_{\odot}$)&($R_{\odot}$)&($M_{\odot}$)&($M_{\odot}$)\\
  \hline\noalign{\smallskip}
V811 Cep&475.8&14.200&5.456&$-$0.220&5.236&0.477&0.162&0.833&0.481&1.618&0.703&0.200\\
(Error)&$\pm$4.3&$-$&$\pm$0.020&$-$&$\pm$0.020&$\pm$0.009&$\pm$0.004&$\pm$0.062&$\pm$0.007&$\pm$0.073&$\pm$0.097& $\pm$0.030\\

  \noalign{\smallskip}\hline
\end{tabular}}
\end{center}
\end{table}

\begin{table}
\begin{center}
\caption[Table5]{ Contact Binary Systems of K-type }\label{VCCH5}
\resizebox{\textwidth}{!}{
 \begin{tabular}{cccccccc}
  \hline\noalign{\smallskip}
Star & Period (d) & $q$ & $f(\%)$ & $T_{1}$(K) & $T_{2}$(K) & $dp/dt(d\ yr^{-1})$ & Reference \\
  \hline\noalign{\smallskip}
AD Cnc & 0.282738 & 0.770 & 8.3 & 5000 & 4790 & $+4.94\times10^{-7}$ & \citet{Qian+etal+2007} \\
LO Com & 0.286361 & 0.404 & 3.2 & 5178 & 4874 & $-1.18\times10^{-7}$ & \citet{Zhang+etal+2016}\\
AU Ser & 0.386479 & 0.692 & 4.0 & 5140 & 4973 & $-1.17\times10^{-7}$ & \citet{Alton+etal}\\
BM UMa & 0.271221 & 0.540 & 17.0& 4600 & 4982 & $-7.49\times10^{-8}$ & \citet{Yang+Wei+Nakajima}\\
VW Cep & 0.278315 & 0.302 & 22.0& 5050 & 5342 & $-1.69\times10^{-7}$ & \citet{Mitnyan+etal}\\
RZ Com & 0.338506 & 0.425 & 20.1& 5000 & 4900 & $+3.97\times10^{-8}$ & \citet{He+Qian}\\
GK Aqr & 0.327413 & 0.435 & 5.1 & 5326 & 4995 & $+2.80\times10^{-7}$ & \citet{Zhang+Pi+Han+2015}\\
MW And & 0.263769 & 0.503 & 40.2& 4579 & 4338 & $+1.97\times10^{-6}$ & \citet{Zubairi+etal}\\
YZ Phe & 0.234726 & 0.380 & 9.7 & 4908 & 4658 & $-2.64\times10^{-8}$ & \citet{Sa+etal}\\
TW Crucis & 0.388144 & 0.670 & 11.0 & 5000 & 5170 & $+2.03\times10^{-6}$ & \citet{Moriarty+D+J+W}\\
V811 Cep & 0.250760 & 0.285 & 33.9 & 5264 & 5289 & $-3.90\times10^{-7}$ & This paper\\
  \noalign{\smallskip}\hline
\end{tabular}}
\end{center}
\tablecomments{0.86\textwidth}{(1) $q=M_{2}/M_{1}$; (2) The subscript 1,2 represent the more massive and less massive star, respectively.}
\end{table}

\begin{figure*}
  \centering
  \begin{minipage}[t]{0.49\textwidth}
  \centering
  \includegraphics[width=\textwidth, angle=0]{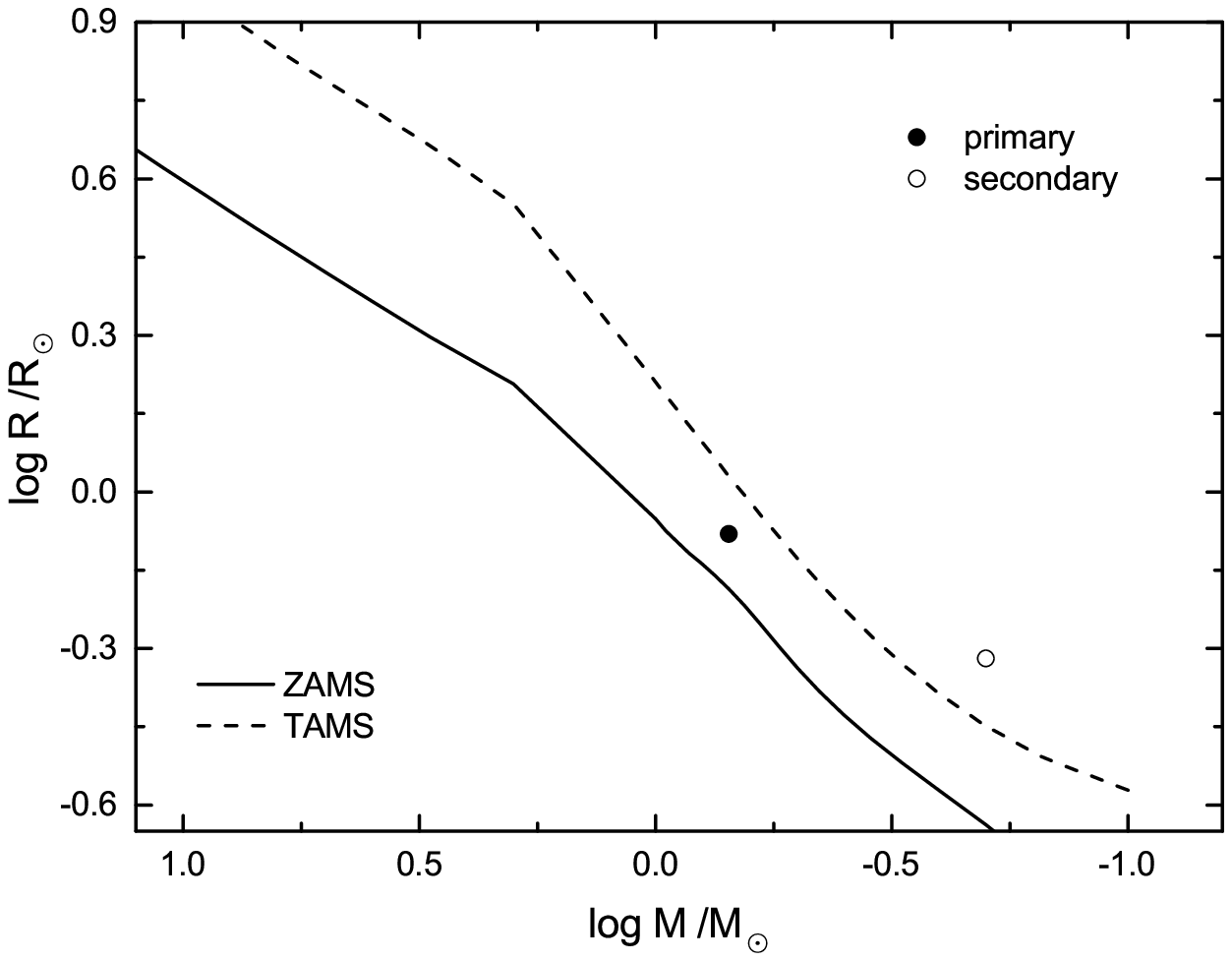}
  \end{minipage}
  \begin{minipage}[t]{0.48\textwidth}
  \centering
  \includegraphics[width=\textwidth, angle=0]{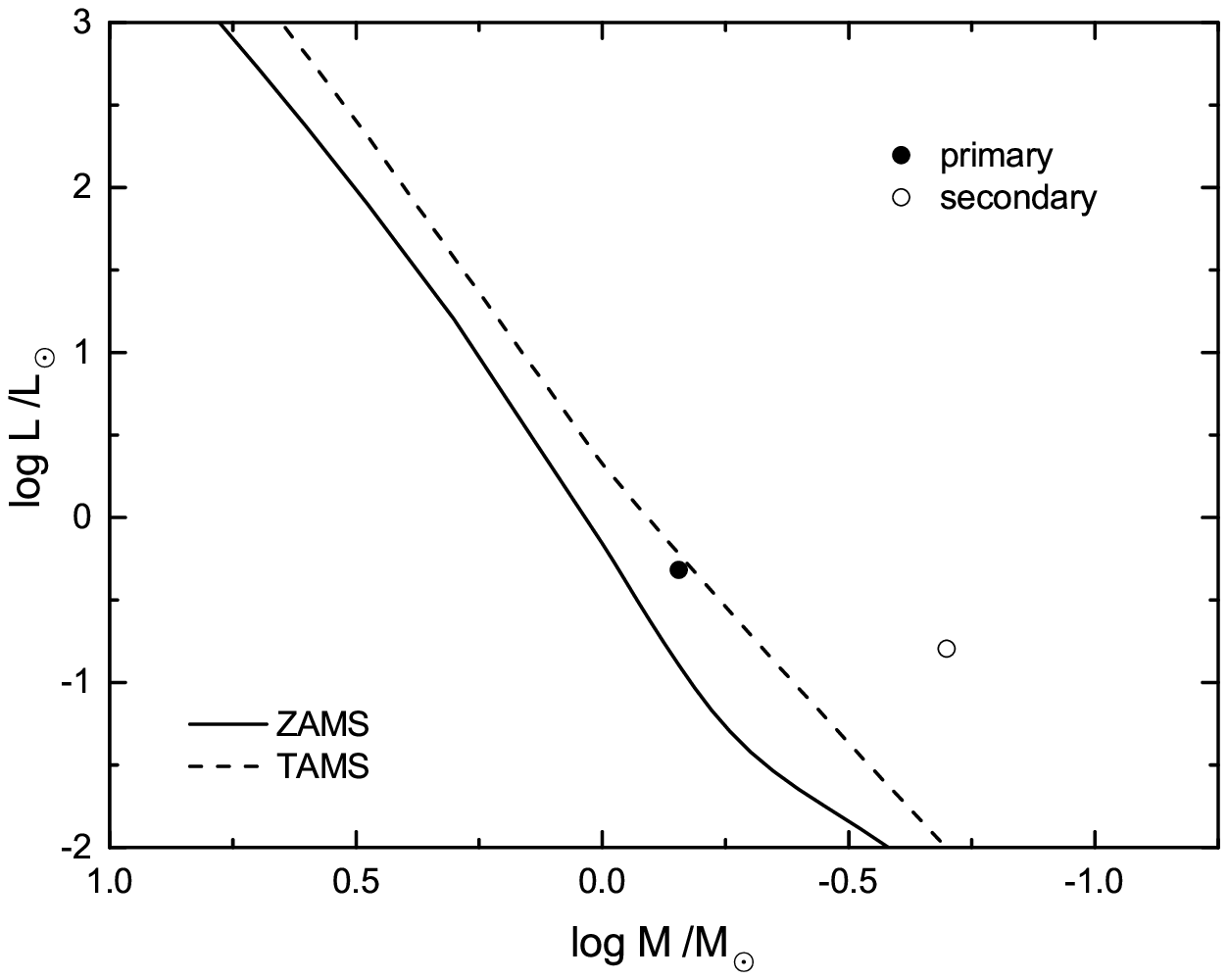}
  \end{minipage}
  \caption{The left figure is the mass-radius diagram, and the right picture is the mass-luminosity diagram. In each figure, the solid circle represents the primary star, and the hollow circle represents the secondary star. The solid and dashed lines in each figure represent ZAMS and TAMS, respectively.}
  \label{Fig6}
  \end{figure*}

The general trend of the O-C curve in the Figure~\ref{Fig2} is a downward parabola, indicating that the period decreases continuously, and the reduction rate was calculated to be $3.90(\pm0.06)\times10^{-7}d \cdot yr^{-1}$. The decrease of period can be explained by the material transfer from the component with more mass to the component with less mass (\citealt{Li+2018+pasp}). The material transfer rate was calculated by the following equation:
\begin{equation}
\label{E3}
\frac{\mathrm d M_1}{\mathrm d t}=\frac{M_1M_2}{3P(M_1-M_2)}\times\frac{\mathrm dP}{\mathrm dt}
\end{equation}
The result of the equation is $1.45\times10^{-7}M_{\odot}\,yr^{-1}$. It has been proved that V811 Cep is a contact binary system of K0-type, and Table~\ref{VCCH5} shows some contact binary systems of K-type. As shown in Table~\ref{VCCH5}, the period change rate of V811Cep is close to that of other systems in the table. In order to further explore the evolution of the two components in this system, the mass-radius diagram and the mass-luminosity diagram are shown in Figure~\ref{Fig6}. Two lines of ``ZAMS'' and ``TAMS'' were constructed by using the binary evolution code of \citet{Hurley+et+al}, in which the former refers to the zero-age main sequence and the latter refers to the terminal-age main sequence. From the two diagrams, we can see that the primary star lies between the two lines, which indicates that the primary star is still in the main sequence stage, while the secondary star is above the TAMS, which indicates that the secondary star has evolved out of the main sequence stage. In Figure~\ref{Fig6}, the secondary component of V811 Cep is over-luminous and over-size. These are thought to be caused by the material transfer from the primary component to the secondary component. The two components of V811 Cep are in the stage of evolution similar to other W-type contact binaries.

This is the first photometric study of V811 Cep so far. The results show that the mass ratio of V811 Cep is 0.285 and the orbital period decreases continuously, which is consistent with the research result of \citet{Qian+S+B+2001}. This result shows that for the W-type contact binaries with $q<0.4$, the orbital periods show a long-term decreasing trend, while for the W-type contact binaries with $q>0.4$, the orbital period shows a long-term increasing trend. The long-term decrease of the period can be explained by the transfer of matter from the primary component to the secondary component. At the same time, the light curve of V811 Cep has obvious O'Connell effect, which can be explained by the existence of a hot spot on the secondary component. On the other hand, with regard to the formation of the hot spot on the secondary component, it is accepted that the hot spot is formed due to the material transfer from the primary component to the secondary component. This coincides with the previous explanation.

\begin{acknowledgements}
We thank the referee very much for the very helpful comments. This work is supported by the Joint Research Fund in Astronomy (No. U1931103) under cooperative agreement between NSFC and Chinese Academy of Sciences (CAS), and by National Natural Science Foundation of China (NSFC) (No. 11703016), and by Young Scholars Program of Shandong University, Weihai (Nos. 20820171006), and by the Open Research Program of Key Laboratory for the Structure and Evolution of Celestial Objects (No. OP201704).
The calculations in this work were carried out at Supercomputing Center of Shandong University, Weihai.

This paper makes use of data from the DR1 of the WASP data (\citealt{Butters+etal+2010}) as provided by the WASP consortium, and the computing and storage facilities at the CERIT Scientific Cloud, reg. no. CZ.1.05/3.2.00/08.0144 which is operated by Masaryk University, Czech Republic.
This work has made use of data from the European Space Agency (ESA) mission Gaia (https://www.cosmos.esa.int/gaia), processed by the Gaia Data Processing and Analysis Consortium (DPAC, https://www.cosmos.esa.int/web/gaia/dpac/consortium). Funding for the DPAC has been provided by national institutions, in particular the institutions participating in the Gaia Multilateral Agreement.

\end{acknowledgements}

\appendix                  

\label{lastpage}

\end{document}